\documentclass[11pt]{article}

\usepackage{geometry}                
\usepackage[small]{caption}
\usepackage{subfig}
\usepackage{graphicx}
\usepackage{amsmath}
\usepackage{amsfonts}
\usepackage{amssymb}
\usepackage{epstopdf} 
\usepackage{mathrsfs}
\usepackage[T1]{fontenc}
\usepackage{feynmf}
\def\7#1#2{\mathop{\null#2}\limits^{#1}}        

\def\beee{\begin{equation}}
\def\eeee{\end{equation}}
\def\dggg{^{\dagger}}

\begin{document}
\bibliographystyle{unsrt}

\begin{center}
\textbf{N-QUANTUM CALCULATION\\ OF THE HYDROGEN ATOM\\ WITH ONE-PHOTON EXCHANGE}\\
\vspace{5mm}
O.W. Greenberg\footnote{email address, owgreen@umd.edu}\\  
Steve Cowen\footnote{email address, scowen@umd.edu}\\
{Center for Fundamental Physics\\
Department of Physics \\
University of Maryland\\
College Park, MD~~20742-4111}\\
University of Maryland Preprint PP-012-023\\
\end{center}

\begin{abstract}      
The N-quantum approach (NQA) to quantum field theory uses the complete and irreducible
set of \textit{in} or \textit{out} fields, including \textit{in} or \textit{out} fields for bound states, as standard building blocks to construct solutions to 
quantum field theories.  In particular, introducing \textit{in} (or \textit{out}) fields for the bound states allows a new way to calculate energy levels and wave functions for the bound states that is both covariant and effectively 3-dimensional. This method is independent of the Bethe-Salpeter equation. In contrast to the Bethe-Salpeter equation, all solutions of the NQA are normalizable and correspond to physical bound states. In this paper we use the NQA in one-loop approximation to calculate states of the relativistic hydrogen atom and analogous two-body systems to illustrate how our new method works. With additional terms in the \textit{in} field
expansion we find systematic corrections beyond the Coulomb interaction.
\end{abstract}

\section{Introduction}
The tour de force experiment of Pohl, et al. \cite{poh} and \cite{Ant} provided new motivation for us to continue work on the NQA of calculating bound state properties.  The discrepancy between the proton structure measured in hydrogen and that measured in muonic hydrogen has three possible causes, (1) new physics, (2) inadequate QED calculations, or (3) incorrect description of the interaction with the proton. To study (2), we are developing a new way to do the QED calculations. In this paper we introduce our new method of calculation, but do not carry the work to the level necessary to resolve the discrepancy found in the Pohl, et al, and Antognini, et al, experiments. Other attempts to resolve the discrepancy can found from the citations to these articles. 
We cite one such calculation \cite{miller} that also examines the second case.  In this paper we present a one-loop calculation of the energy levels and wave functions of ordinary hydrogen and muonic hydrogen. The NQA, based on Haag's expansion~\cite{haa} of interacting fields in terms of asymptotic fields, also can be applied to other two-body systems, such as the  $(e \bar{\mu})$ and $(\mu \bar{\mu})$ systems.

Haag's original expansion did not take account of bound states. We added \textit{in} (or \textit{out}) fields for each of the bound states that we take as stable in our approximate treatment. For the case of the hydrogen atom we add an 
\textit{in} (or \textit{out}) field for every state of the hydrogen atom.

\section{Goals of this paper}

Our main goal in this paper is to develop a new method of calculation of the energy levels and wave functions of relativistic bound states. This method has been described previously~\cite{nqa,gg,ray,the}, but it has not been developed sufficiently to account for high-order radiative and recoil effects that are relevant to the analysis of high-precision spectroscopic measurements, such as those in ordinary and muonic hydrogen. Among the advantages of this method for (2-body)  bound states such as the hydrogen atom are the introduction of both masses as independent parameters, rather than via the reduced mass. Since we are interested in muonic hydrogen as well as ordinary hydrogen, taking the proton mass as an independent parameter is important. We also solve integral equations that incorporate radiative and recoil effects without perturbation theory. References to other applications of the NQA are in \cite{vir}. Of particular relevance to our present paper is \cite{gg}, which is the first paper to give the spectator equation for a two-body bound state. There is related work by K\"all\'en, \cite{kal}, Yang and Feldman \cite{yan} and, for bound states, by Gross \cite{gro}.
 
In this paper we present a one-loop approximation for the relativistic bound state that reduces to the Dirac equation for large proton mass. As stated above, we introduce the proton mass as an independent parameter. This is relevant in the case of muonic hydrogen where the mass ratio $m_{\mu}/M_p \approx 1/10$ rather than the ratio $m_e/M_p \approx 1/2000$ for ordinary hydrogen. We use the Coulomb potential as the binding mechanism and ignore magnetic interactions and renormalization effects in this paper.  We do not calculate energy
levels in high accuracy in the present paper. Rather, we describe our new method of calculation. In later papers, in addition to our systematic development of the NQA, we will give a unified calculation that includes all quantum electrodynamic corrections to the hydrogen spectrum up to a given order, rather than adding various corrections piece by piece. In our next paper we will pay particular attention to any differences between our results and the usual calculations to see whether this new method resolves the muonic hydrogen anomaly concerning the proton charge radius as inferred from measurements of the Lamb shift in ordinary and muonic hydrogen. We will also find a set of coupled integral equations that include all terms up to a relevant order, rather that adding corrections term by term. 

\section{Asymptotic fields and the Haag expansion}

The \textit{in} (\textit{out}) fields have free field commutators, obey free equations of motion,
and the different \textit{in} (\textit{out}) fields commute or anticommute with each other everywhere in spacetime.  Each of these sets of asymptotic fields by themselves is completely
known once the masses, spins, and quantum numbers of the fields in a given set are 
given.  Thus either set serves as a collection of standard building blocks to construct solutions of the operator equations of motion. For the present paper we ignore the difficulty that the asymptotic limits for the charged fields do not exist. In a recent paper~\cite{gc} we found modified charged fields for which the asymptotic limits do exist; however we do not use the modified charged fields here. 
 
To solve the equations of motion using the Haag expansion, expand the fields that
appear in the Hamiltonian or Lagrangian in normal-ordered series of \textit{in} (\textit{out}) 
fields.  To determine the $c$-number amplitudes (the Haag amplitudes) 
that are the coefficients of the normal-ordered
terms, insert the expansion in the operator equations of motion, 
renormal-order, and equate the coefficients of corresponding (linearly independent)
normal-ordered terms.  The relevant Haag 
amplitudes are the wave functions of the bound states.  

The method based on
the Haag expansion is entirely independent of the Bethe-Salpeter equation \cite{sal}. 
In contrast to the Bethe-Salpeter approach, in the NQA there are no spurious solutions and no negative norm amplitudes.  The NQA can be used for bound states in relativistic theories
even though the amplitudes depend only on the same number of kinematic
variables as nonrelativistic wave functions.  In particular, there are no
relative times in the relativistic version of the NQA. With all terms in the $in$ field expansions allowed by conservation laws, the Haag expansion should be equivalent to the interacting field theory. This results in an infinite set of coupled equations. To get a tractable set of equations, we terminate the Haag expansions, keeping a finite set of terms for each interacting field. For quantum electrodynamics, the case relevant here, the smallness of $\alpha$ provides a rationale to terminate the series; each vertex has a factor of $\sqrt{\alpha}$. For this paper we restrict to the simplest terms in the Haag expansion that give an equation for the bound state.

We need not construct the $in$ field for the bound state. We assume possible bound states and introduce $in$ fields, characterized by their mass, spin, and other quantum numbers, for each. The equations of motion for the interacting fields give equations for the bound state amplitudes; if there is a solution for a given bound state amplitude, then the corresponding bound state exists. We take ``bound state amplitude'' as a synonym for ``wave function.'' 
 
\section{Relativistic model of the hydrogen atom}

The fundamental fields are the electron, $e_{\alpha}(x)$, muon, $\mu_{\alpha}(x)$, proton, $p_{\alpha}(x)$, 
and photon vector potential, $A_{\mu}(x)$  fields. These fields obey the operator equations of motion (for this paper we drop renormalization counter terms),
\begin{gather} 
(i\not\! \partial  -m) e(x) =\frac{e}{2} [\not\!\!A(x),e(x)]_+ , \\
(i\not \!\partial  -m_{\mu})\mu(x) =\frac{e}{2} [\not\!\!A(x),\mu(x)]_+ , \\
(i \not \!\partial -M) p(x) =-\frac{e}{2} [\not\!\!A(x),p(x)]_+ ,  \\
\partial^{\mu} \partial \cdot A- \partial \cdot \partial A^{\mu}=
\frac{e}{2}([\bar{e}(x), \gamma^{\mu} e(x)]_-+[\bar{\mu}(x), \gamma^{\mu} \mu(x)]_- - [\bar{p}(x), \gamma^{\mu} p(x)]_-) \label{eq:photeq}
\end{gather}
where  Eq.\eqref{eq:photeq} follows from 
\begin{align}
\partial_{\nu}  F^{\mu \nu}(x) = \frac{e}{2}([\bar{e}(x), \gamma^{\mu} e(x)]_-
+ [\bar{\mu}(x), \gamma^{\mu} \mu(x)]_-- [\bar{p}(x), \gamma^{\mu} p(x)]_-)
\end{align}
and $F^{\mu \nu}(x) = \partial^{\mu} A^{\nu}(x)-  \partial^{\mu} A^{\nu}(x)$.
We choose the masses of the electron, muon, proton, and hydrogen atom $in$ states $i$ as $m$,
$m_{\mu}$, $M$, and $M_i$, respectively. The equations are symmetric under 
$e \leftrightarrow \mu,~m \leftrightarrow m_{\mu}$.

We use the Haag expansion to expand the interacting fields appearing in the equations of motion in terms of \textit{in} fields.  We truncate the series, keeping the first term involving the hydrogen bound state \textit{in} fields, $h_i^{in}$,
\begin{align}
e(x)&=e^{(in)}(x)+\sum_i \int d^3 y d^3 z :\bar{p}^{(in)}(y) f_{\bar{p}h,i}(x-y, x-z)
i\stackrel{\longleftrightarrow}{\frac{\partial}{\partial z^0}}h_i^{(in)}(z): \\
\bar{e}(x)&=\bar{e}^{(in)}(x)+\sum_i\int d^3 y d^3 z :h_i^{(in)\dagger}(z)
i\stackrel{\longleftrightarrow}{\frac{\partial}{\partial z^0}}\bar{f}_{\bar{p}h}(x-y, x-z) p^{(in)}(y): \\
p(x)&=p^{(in)}(x)+\sum_i \int d^3 y d^3 z f_{\bar{e}h}(x-y, x-z) :\bar{e}^{(in)}(y) i\stackrel{\longleftrightarrow}{\frac{\partial}{\partial z^0}}h_i^{(in)}(z): \\
\bar{p}(x)&= \bar{p}^{(in)}(x)+ \sum_i \int d^3 y d^3 z :h_i^{(in)\dagger}(z)i\stackrel{\longleftrightarrow}{\frac{\partial}{\partial z^0}}e^{(in)}(y):\bar{f}_{\bar{e}h}(x-y, x-z) \\
A^{\mu}(x)&=A^{(in)\mu}(x) + \int d^3y d^3z 
[:\bar{p}^{(in)}(y) f^{\mu}_{\bar{p}p}(x-y, x-z) p^{(in)}(z):\notag \\
&\;\;\;\;\;\;\;\;\;\;\;\;\;\;\;\;\;\;\;\;\;\;\;\;\;\;\;\;\;\;\;\;\;\;\;\;\;\;+
:\bar{e}^{(in)}(y) f^{\mu}_{\bar{e}e}(x-y, x-z) e^{(in)}(z):]
\end{align}
where the $\sum_i$ is the sum over the various hydrogen states which, for simplicity, we took as scalar,  
\begin{align}
\bar{f}_{\bar{p}h}(x, y)&=\gamma^0f^{\dagger}_{\bar{p}h}(x, y)\gamma^0 \\
\bar{f}_{\bar{e}h}(x, y)&=\gamma^0f^{\dagger}_{\bar{e}h}(x, y)\gamma^0.
\end{align}
and we label each amplitude by the \textit{in} fields in each term in the expansion of the interacting fields, and keep this label for the terms in the adjoints of the interacting fields.

We used translation invariance to write these forms of the expansions. Lorentz covariance gives the transformation properties of the amplitudes:
\begin{align}
S(\Lambda) f_{\bar{p}h}(x, y)S(\Lambda)^{-1}=f_{\bar{p}h}(S(\Lambda) x, S(\Lambda) y).
\end{align}
We choose spectroscopic notation for the states of the hydrogen atom that is adapted to treating the proton spin on the same basis as the electron spin. We use $F$, $L$, $S$ for the total angular
momentum (an exact quantum number), the orbital angular momentum, and the lepton-proton spin, respectively. With the principal quantum number, $n$, we label states as $nL_S^F$. (Our choice differs from the usual choice that couples the orbital angular momentum, $L$, to the electron spin, $S_e$, then couples 
$J=L+S_e$ to the proton spin to get $F$, and labels the states as $nL_J^F$.)

Because the contractions, $\langle 0|e^{(in)}(x) \bar{e}^{(in)}(y)|0 \rangle$, etc. are
simpler in momentum space than in position space, we continue our analysis in momentum space.
To go into momentum space, we use 
\begin{align}
e(x)=\int d^4q  e(q) \exp(-i q \cdot x)
\end{align}
and analogous formulas for the other fields. We leave tildes off the Fourier-transformed fields. The equations in momentum space are
\begin{align} 
(\not \! q-m) e(q)&=\frac{e}{2} \int d^4k [\not\!\! A(k), e(q-k)]_+ \\
(\not \! q-m_{\mu}) \mu(q)&=\frac{e}{2} \int d^4k [\not\!\! A(k), \mu(q-k)]_+  \\
(\not \! p-M) p( p)&=-\frac{e}{2} \int d^4k [\not\!\! A(k), p(p-k)]_+  \\
-k^\mu k\cdot A(k)+k^2 A^\mu (k)&=\frac{e}{2}\int d^4q' ([\bar{e}(q'),\gamma^\mu e(k-q')]\\
&\;\;\;\;+[\bar{\mu}(q'),\gamma^\mu \mu(k-q')]-[\bar{p}(q'),\gamma^\mu p(k-q')]) 
\end{align}
To avoid subscripts we use $h$ for ordinary hydrogen (electronic hydrogen) and $H$ for muonic hydrogen. For this one-loop approximation we choose the Coulomb gauge. We expand the interacting fields in normal-ordered products of $in$ fields. For the one-loop approximation to the amplitude in which $e \sim :\bar{p}^{(in)}h^{(in)}:$, we keep terms with up to three $in$ fields in the Haag expansions for $e$ and $A$ and one contraction. These terms are 
$:\bar{p}\overbrace{p::\bar{p}}h:$, $:\overbrace{A::A}\bar{p}h:$, $\bar{p}\overbrace{p::\bar{p}}Ah:$,  where the overbraces stand for contractions. The term from $:h\overbrace{\bar{h}::h}\bar{p}:$ is much higher order because the hydrogen atom has zero charge.  The Haag expansion for the electron field is
\begin{align}
e(q) &=e^{in}(q)+\sum_j \int d^4p d^4b \delta(p+q-b)f_{\bar{p}h_j}(p,b)
:\bar{p}^{in}(p)h_j^{in}(b): \notag \\
& \;\;\;\;+\int d^4p d^4b \delta(p+q+k-b)
f^{\mu}_{A\bar{p}h_j}(p,k,b) :A_{\mu}(-k)\bar{p}^{in}(p)h_j^{in}(b):
\\
\bar{e}(q) &=\bar{e}^{in}(q)+\sum_j \int d^4p d^4b \delta(p+q-b):h^{in \,\dagger}(b)p^{in}(p):
\bar{f}_{\bar{p}h_j}(p,b): \notag
\\
&\;\;\;\; +\int d^4p d^4b \delta(p+q+k-b):h^{in \, \dagger}(b)p^{in}(p)
A_{\mu}(k):\bar{f}_{A\bar{p}h_j}(p,b): ,
\end{align}
where $\bar{f}_{\bar{p}h_i}(p,b)=\gamma^{0T}f^{\dagger}_{\bar{p}h_i}(p,b)\gamma^0$.\\
We chose this parametrization so that
\begin{align}
(\bar{e}(q)\bar{p}(p)|0\rangle,h_i\dggg(b)|0\rangle) = \delta(q+p-b) \bar{f}_{\bar{p}h_i}(p,b)(\not\! p+M)\theta(p^0) \delta(p^2-M^2)\theta(b^0)\delta(b^2-M_i^2).
\end{align}
There are analogous expressions for the muon and proton fields.
For the photon field,
\begin{align}
A^{\mu}(k)&=A^{\mu~in}(k)+\int d^4p d^4p^{\prime}\delta(k-p-p^{\prime})
[:\bar{p}^{in} (p)f^{\mu}_{p p^{\prime}}(p,p^{\prime}) p^{ in}(p^{\prime}): \notag \\
& \;\;\;\;\;\;\;\;\;\;\;\;\;\;\;\;\;\;\;\;\;\;\;\;-:\bar{e}^{in} ( p)f^{\mu}_{e e^{\prime}}(p,p^{\prime}) e^{ in}(p^{\prime}): -:\bar{\mu}^{in} ( p)f^{\mu}_{\mu \mu^{\prime}}(p,p^{\prime}) \mu^{ in}(p^{\prime}):].
\end{align}
After re-normal-ordering, we find the one-loop equations for the two main amplitudes for any state of the hydrogen atom,
\begin{align} \label{eq:bse1}
(\displaystyle {\not} b - {\not} p-m)f_e(p,b)  &=\frac{e^2}{2 (2\pi)^3}
\int\frac{d^3p'}{2 E_{p'}} \gamma^{\mu}\frac{f_e(p',b)}{(p-p')^2}(\gamma_{\mu})^T(\displaystyle{\not} p+M)^T\notag \\
&\;\;\;\;   -\frac{e^2}{2 (2\pi)^3}
\int\frac{d^3p'}{2 e_{p'}} \gamma^{\mu}\frac{f_p(p',b)^T}{(b-p'-p)^2}(\gamma_{\mu})^T(\displaystyle{\not} p+M)^T \\  \label{eq:bse2}
(\displaystyle {\not} b - {\not} q-M)f_p(q,b)  &=\frac{e^2}{2 (2\pi)^3}
\int\frac{d^3q'}{2 e_{q'}} \gamma^{\mu}\frac{f_p(q',b)}{(q-q')^2}(\gamma_{\mu})^T(\displaystyle{\not} q+M)^T\notag \\
&\;\;\;\; -\frac{e^2}{2 (2\pi)^3}
\int\frac{d^3q'}{2 E_{q'}} \gamma^{\mu}\frac{f_e(q',b)^T}{(b-q'-q)^2}(\gamma_{\mu})^T(\displaystyle{\not} q+m)^T. 
\end{align}
where $f_e(p,b)\equiv f_{\bar{p} h}(p,b) (\displaystyle{\not} p+M)^T$, $f_p(q,b)\equiv f_{\bar{e} h}(q,b) (\displaystyle{\not} q+m)^T$, $E_p=\sqrt{\mathbf{p}^2+M^2}$, $e_q=\sqrt{\mathbf{q}^2+m^2}$, $p$ is the energy-momentum of the on-shell proton, and $q=b-p$ is the energy-momentum of the off-shell electron.  Note that, by construction, $f_e(p,b)$ obeys the subsidiary condition
$f_e(p,b)(\displaystyle{\not} p-M)^T=0$ and $f_p(q,b)$ obeys $f_p(q,b)(\displaystyle{\not} q-m)^T=0$.  Unlike the Bethe-Salpeter approach, we have arrived at a pair of coupled equations that describe the bound state.  They are explicitly symmetric under subscript $e\leftrightarrow p$ and mass $m\leftrightarrow M$ interchange.  These two equations differ from those found in \cite{Lepage} and \cite{Faustov} where Bethe-Salpeter equations with one on-shell particle are found, but we will show that in a certain approximation they reduce to their Bethe-Salpeter counterparts.  As far as we know, the exact properties of Eqs. (\ref{eq:bse1}) and (\ref{eq:bse2}) are unexplored.  

\begin{figure}[t] 
\begin{center}
\subfloat[]{\label{fig:a}
\begin{fmffile}{fd}
\begin{fmfgraph*}(100,100) \fmfpen{thin}
\fmfbottomn{i}{2} \fmftop{o1} 
\fmf{fermion,width=25,label=$b-p$}{i1,v1}
\fmf{fermion,width=25}{v1,v3}
\fmf{photon,width=25,tension=0}{v1,v2}
\fmf{fermion,label=$p$}{i2,v2}
\fmf{fermion}{v2,v3}
\fmf{dashes,label=$b$}{v3,o1}
\fmfdot{v1,v2}
\fmfv{decor.shape=circle,decor.filled=empty,
decor.size=5thick}{v3}
\end{fmfgraph*}
\end{fmffile}}
\;\;\;\;\;\;\;\;\;\;\;\;
\subfloat[]{\label{fig:b}
\begin{fmffile}{diagram2}
\begin{fmfgraph*}(100,100) \fmfpen{thin}
\fmfbottomn{i}{2} \fmftop{o1} 
\fmf{fermion,width=25,label=$b-p$}{i1,v1}
\fmf{photon,width=25,tension=0}{v1,v2}
\fmf{fermion}{v1,v3}
\fmf{fermion,label=$p$}{i2,v2}
\fmf{fermion,width=25}{v2,v3}
\fmf{dashes,label=$b$}{v3,o1}
\fmfdot{v1,v2}
\fmfv{decor.shape=circle,decor.filled=empty,
decor.size=5thick}{v3}
\end{fmfgraph*}
\end{fmffile}}

\end{center}
\caption{Graphs for the right hand side of the electron equation of motion.  Heavy lines are off-shell and light lines are on-shell.  The dashed line represents the bound state (hydrogen atom).  The empty circle represents the amplitude $f_e$ in (a) and $f_p$ in (b).  The left fermion line is the electron and the right line is the proton.  Similar graphs exist for the proton equation.}
\end{figure}
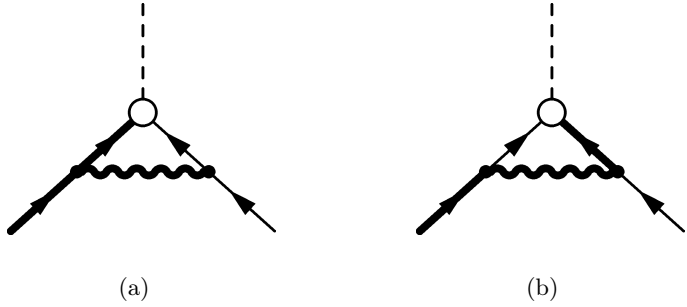

The terms on the right hand side of Eqs. (\ref{eq:bse1}) and (\ref{eq:bse2}) are expressed in diagrammatic form in figures \ref{fig:a} and \ref{fig:b}.  Heavy lines are off-shell and light lines are on-shell.  Point vertices represent the substitution of an off-shell interacting field in terms of other interacting fields via the relevant equation of motion.  These are the fundamental QED vertices.  Circles indicate the use of the Haag expansion to express off-shell interacting fields in terms of \emph{in} fields with a Haag amplitude coefficient.  We will use these diagrams in section 9 to show how higher order corrections are calculated.

In a covariant gauge these equations are explicitly Lorentz covariant for an arbitrary state of motion of the hydrogen atom. The general expansions have the support of the $in$ fields on both mass shells. Here we keep only the mass shell that does not lead to extra pairs of particles. 

\section{Normalization of the wave functions}

The asymptotic fields diagonalize conserved observables such as the 
Hamiltonian, the momentum operators and various charges.  We can represent any conserved quantity $\mathcal{O}$ in terms of either the interacting fields or the asymptotic fields (either the $in$ or $out$ fields),
\begin{align}
\sum_{i}\mathcal{O}[\xi_i]=\sum_{i}\mathcal{O}[\psi^{in}_i],
\end{align}
where $\xi_i$ in an interacting field and $\psi_i^{in}$ is an asymptotic field, including the $in$ fields for bound states. We do not include
weak interactions in our analysis of the hydrogen atom; thus both the number of electrons and the number of muons are conserved. We use the number of electrons, $N_e$, to find the normalization condition for the hydrogen wave function, $f_{\bar{p}h}(p,b)$, in which the electron is off-shell. 
The only interacting field that carries electron number is the electron field,
\begin{align}
N_e=\int e^{\dagger}(x) e(x) d^3 x.
\end{align}
The contribution to the electron number from the hydrogen atom in a given state comes from the terms in $N_e$ that are bilinear in the hydrogen atom $in$ field in that state. From the $in$ field expansion of $e(x)$ we find the orthonormalization condition
\begin{align}
\int \frac{M d^3p}{E_p}\text{Tr}[\bar{f}_{\bar{p}H_{j'}}(p,b')\gamma^0 f_{\bar{p}H_{j}}(p,b)]
=\delta_{j' j}2E_b \delta(\mathbf{b}'-\mathbf{b}).
\end{align}

\section{Interchange of the on-shell and off-shell particles}

The equal-time anticommutators relate the Haag amplitudes with the lepton off-shell to those with the proton off-shell. These relations follow from the vanishing of the coefficients of each
(linearly independent) normal-ordered product of $in$ fields in the equal-time anticommutators.  Most of the relations
involve Haag amplitudes for terms with higher degree normal-ordered products
than we have considered here; however for the equal-time anticommutator $[e,p]_+=0$
there is an approximate relation that involves only terms that we considered here,
\beee \label{eq:sym}
[f_{\bar{e}H,i} (q,b)(\frac{\displaystyle{\not} q + m}{2e_q})^T]_{\beta \alpha} + [f_{\bar{p}H,i} (p,b)(\frac{\displaystyle{\not} p + M}{2E_p})^T]_{ \alpha \beta}=0,
\eeee
with the constraint $p+q=b$.
Thus, the Haag amplitude with the lepton off-shell is simply related to that with
the proton off-shell.   The two amplitudes determine each other 
uniquely.

Using Eq. (\ref{eq:sym}), we can simplify Eq. (\ref{eq:bse1}) to
\begin{align} \label{eq:bse3}
(\displaystyle {\not} b - {\not} p-m)f_e(p,b)  &=\frac{e^2}{(2\pi)^3}
\int\frac{d^3p'}{2 E_{p'}} \gamma^{\mu}\frac{f_e(p',b)}{(p-p')^2}(\gamma_{\mu})^T(\displaystyle{\not} p+M)^T,
\end{align}
which matches the Bethe-Salpeter equations of \cite{Lepage} and \cite{Faustov}.
For the present paper, we choose the hydrogen atom at rest, $b=(M_{H_i}, \mathbf{0})$, which explicitly breaks Lorentz covariance to rotation covariance. Keeping the main mass shell, and dropping the magnetic interaction terms, we have
\begin{align} \label{eq:bse}
(\gamma^0 M_{H_i}-\displaystyle{\not} p-m)f_e(\mathbf{p})  &=-
\frac{e^2}{(2\pi)^3}\int\frac{d^3p'}{2 E_{p'}}  \gamma^0 
\frac{f_e(\mathbf{p}')}{|\mathbf{p}-\mathbf{p'}|^2}(\gamma^0)^T(\displaystyle{\not} p+M)^T,
\end{align}
where $f_e(\mathbf{p})\equiv f_e(p;M_{H_i},\mathbf{0})$.

\section{Solution to bound state equation}
The purpose of the following sections is to find a method for solving Eq. (\ref{eq:bse}) which can be extended to solve Eqs. (\ref{eq:bse1}) and (\ref{eq:bse2}).  For the sake of simplicity in this work, we will put off solving our more complicated coupled equations in a future paper.  We acknowledge that we are solving an equation that has already been studied extensively in the literature, but our purpose is to develop the NQA framework for high precision calculations.  We therefore take an approach that differs from the typical perturbative method.   The methods for finding higher order corrections will be discussed in section 9.

In this section, we focus on binding due to the Coulomb interaction.  We chose Coulomb gauge to simplify our calculations and to allow comparison with the usual solution of the Dirac equation for the hydrogen atom.  To keep the notation general for any two-particle system, we label the constituents $m_1$ and $m_2$ and the bound state $m_b$ in this section.  We solve the bound-state equation
\begin{align} \label{eq:bsenew}
(\gamma^0 m_b-\displaystyle{\not} p-m_1)f_e(\mathbf{p})  &=\frac{e^2}{(2\pi)^3}\int \frac{d^3p'}{2 E^{(2)}_{p'}}\gamma^0 V(\mathbf{p,p'})  f_e(\mathbf{p}')(\gamma^0)^T(\displaystyle{\not} p+m_2)^T      
\end{align}
where $m_b=m_1 + m_2 +\epsilon_i$,  $m_b$ is the energy of the hydrogen state, $\epsilon_i <0$ is the binding energy of the atom, $V(\mathbf{p,p'})=- 1/|\mathbf{p-p'}|^2$,  and 
$E^{(i)}_p=\sqrt{p^2+m_i^2}$.  A similar equation is solved in \cite{ray, the} in the 
non-relativistic limit.  We solve the equation numerically without taking a non-relativistic limit.  

We can think of $m_2$ as the mass of the proton and $m_1$ as the mass of the lepton, but the N-quantum equations that give Eq.(\ref{eq:bse}) are symmetric under $m_1 \leftrightarrow m_2$ together with $e \leftrightarrow -e$ and our calculations reflect this.  

Before solving this equation, we show that it reduces to the expected Dirac equation in the large $m_2$ limit.  The factor $(\displaystyle{\not} p+m_2)^T/2E_{p'} \rightarrow (1+\gamma^0)/2$ in the potential in Eq.(\ref{eq:bse}) reduces the $4 \times 4$ system of equations to a $4 \times 2$ system with the usual Coulomb potential. From 
Eq.(\ref{eq:bse}), using $q=b-p$, we find
\begin{equation}
(E \gamma^0 - \mathbf{\gamma} \cdot \mathbf{q}  -m_1 -\gamma^0 V)f_e=0,
\end{equation}
where $V$ is the Coulomb potential. Because this equation comes from a covariant formulation, we have to multiply from the left by $\gamma^0 = \beta$ to get the usual form of the Dirac equation for the hydrogen atom,
\begin{equation}
(\mathbf{\alpha} \cdot \mathbf{q}  +m_1 \beta +V)f_e =Ef_e,
\end{equation} 
using $\gamma^0=\beta$ and $\gamma^0 \mathbf{\gamma^i}=\mathbf{\alpha^i}$.

\subsection{Form of the matrix}
To solve this equation, we break the $4\times 4$ matrix down into four $2\times 2$ matrices:
\begin{equation}
 f_e(\mathbf{p}) = \left(  \begin{array}{cc}
  A(\mathbf{p}) & B(\mathbf{p})  \\
  C(\mathbf{p}) & D(\mathbf{p})  \\ \end{array}  \right).
\end{equation}
Next we introduce the partial wave expansion of the operators and the amplitude.
Each of these $2\times 2$ matrices can be written as a product of a spin-angle part and a radial function.  For example, for a specific eigenstate we can write
\begin{align}
A(\mathbf{p})=\mathscr{Y}^{F m_F}_{LS}(\Omega) g_{L}( p),
\end{align}
where $\mathscr{Y}^{F m_F}_{LS}(\Omega)$ is the spin-angle function, $g_{L}( p)$ is a radial function and $p=|\mathbf{p}|$.  The most general solution is a sum over all possible eigenstates.  The spin-angle function is given by
\begin{align}
\mathscr{Y}^{F m_F}_{LS}(\Omega)&=\sum_{m_L}<LS;m_L\, m_F-m_L | F m_F>\boldsymbol{\phi}_{S\, m_F-m_L} Y_{L m_L}(\theta, \phi).
\end{align}
where $\boldsymbol{\phi}_{S\, m_S}$ is the total spin state of the constituents, $Y_{L m_L}(\theta, \phi)$ is a spherical harmonic, and $<LS;m_L\, m_F-m_L | F m_F>$ is a Clebsch-Gordan coefficient.  The spin state can be either a singlet or a triplet.  We express these in terms of 2-component Pauli spinors:
\begin{equation}
 \boldsymbol{\phi}_{00}=\frac{1}{\sqrt{2}}(\psi(\uparrow) \otimes \chi(\downarrow)-\psi(\downarrow) \otimes \chi(\uparrow)) = \frac{1}{\sqrt{2}}\left(  \begin{array}{cc}
  0 & 1  \\
  -1 & 0  \\ \end{array}  \right)
\end{equation}
\begin{equation}
\boldsymbol{\phi}_{11}=\psi(\uparrow) \otimes \chi(\uparrow) = \left(  \begin{array}{cc}
  1 & 0  \\
  0 & 0  \\ \end{array}  \right)
\end{equation}
\begin{equation}
\boldsymbol{\phi}_{1-1}=\psi(\downarrow) \otimes \chi(\downarrow) = \left(  \begin{array}{cc}
  0 & 0  \\
  0 & 1  \\ \end{array}  \right)
\end{equation}
\begin{equation}
 \boldsymbol{\phi}_{10}=\frac{1}{\sqrt{2}}(\psi(\uparrow) \otimes \chi(\downarrow)+\psi(\downarrow) \otimes \chi(\uparrow)) = \frac{1}{\sqrt{2}} \left(  \begin{array}{cc}
  0 & 1  \\
  1 & 0  \\ \end{array}  \right).
\end{equation}

We expect our matrix wave function to be analogous to the direct product of an electron and a proton spinor,
\begin{equation}
\Phi\equiv\Psi_e \otimes \Psi_p^T =  \left(  \begin{array}{cc}
  \psi_e\otimes \psi_p^T & \psi_e \otimes (\boldsymbol{\sigma} \cdot \mathbf{p} \psi_p)^T    \\
  (\boldsymbol{\sigma} \cdot \mathbf{p} \psi_e)\otimes \psi_p & (\boldsymbol{\sigma} \cdot \mathbf{p} \psi_e) \otimes (\boldsymbol{\sigma} \cdot \mathbf{p} \psi_p)^T  \\ \end{array}  \right).
\end{equation}
where $\Psi_e$ and $\Psi_p$ are free Dirac spinors for the electron and proton respectively, and $\psi_e$ and $\psi_p$ are their upper components.  Our wave function must also satisfy the auxiliary condition
\begin{align} \label{eq:aux}
f_e(\mathbf{p})(\displaystyle{\not} p-m_2)^T=0.
\end{align}
With these two things in mind, we use the form
\begin{equation} \label{eq:wave}
 f_e(\mathbf{p})= \left(  \begin{array}{cc}
  \mathscr{Y}_{LS}^{F m_F} g_{L}( p)\mathbf{1} &  S( p) \mathscr{Y}_{LS}^{F m_F}(\boldsymbol{\sigma}\cdot \hat{\mathbf{p}})^T g_{L}( p)  \\
   \boldsymbol{\sigma}\cdot \hat{\mathbf{p}}\mathscr{Y}_{LS}^{F m_F}  h_{L}( p) & S( p) \boldsymbol{\sigma}\cdot \hat{\mathbf{p}}\mathscr{Y}_{LS}^{F m_F}(\boldsymbol{\sigma}\cdot \hat{\mathbf{p}})^T h_L( p) \\ \end{array}  \right). 
\end{equation}
where $S(p)=p/(E^{(2)}_p+m_2)$, and $\hat{\mathbf{p}}$ is the unit vector in the direction of $\mathbf{p}$.  We constructed this wave function to satisfy 
Eq.\eqref{eq:aux}.
This wave function is also a parity eigenstate, 
\begin{align} 
 \gamma^0 f_e(-\mathbf{p})\gamma^{0 T}= (-1)^L f_e(\mathbf{p}). 
\end{align}

\subsection{The coupled radial integral equations}
Using Eq.\eqref{eq:wave} and the LHS of Eq.\eqref{eq:bse} gives
\begin{equation}
LHS= \left(  \begin{array}{cc}
 L_{11} & L_{12} \\
   L_{21} & L_{22} \\ \end{array}  \right) 
\end{equation}
where
\begin{align*}
L_{11}&=[(m_b-E^{(2)}_p-m_1) g_{L}( p)   +p h_{L}( p)]\mathscr{Y}_{LS}^{F m_F} \\
L_{12}&=  S( p)[(m_b-E^{(2)}_p-m_1) g_{L}( p)   +p  h_{L}( p)]\mathscr{Y}_{LS}^{F m_F} (\boldsymbol{\sigma}\cdot \hat{\mathbf{p}})^T   \\
L_{21}&= -[p g_{L}( p)+(m_b+m_1-E^{(2)}_p)  h_L( p)]\boldsymbol{\sigma}\cdot \hat{\mathbf{p}}\mathscr{Y}_{LS}^{F m_F}   \\
L_{22}&=  -S( p)[ p  g_{L}( p)+(m_b+m_1-E^{(2)}_p)  h_L( p)]\boldsymbol{\sigma}\cdot \hat{\mathbf{ p}}\mathscr{Y}_{LS}^{F m_F}(\boldsymbol{\sigma}\cdot \hat{\mathbf{p}})^T.
\end{align*}
The RHS becomes
\begin{equation} 
RHS= \left(  \begin{array}{cc}
 R_{11} & R_{12} \\
   R_{21} & R_{22} \\ \end{array}  \right) 
\end{equation}
where
\begin{align*}
R_{11}&=\int d^3p' \frac{V(\mathbf{p,p'})}{2 E^{(2)}_{p'}}[(E^{(2)}_p+m_2)\mathscr{Y'}_{LS}^{F m_F}+S(p') p \mathscr{Y'}_{LS}^{F m_F}(\boldsymbol{\sigma}\cdot \hat{\mathbf{p'}})^T (\boldsymbol{\sigma}\cdot \hat{\mathbf{p}})^T] g_L(p')  \\
R_{12}&= \int d^3p' \frac{V(\mathbf{p,p'})}{2 E^{(2)}_{p'}} S( p)[(E^{(2)}_p+m_2)\mathscr{Y'}_{LS}^{F m_F}(\boldsymbol{\sigma}\cdot \hat{\mathbf{p}})^T+S(p') p \mathscr{Y'}_{LS}^{F m_F}(\boldsymbol{\sigma}\cdot \hat{\mathbf{p'}})^T ]g_L(p')  \\
R_{21}&=-\int d^3p' \frac{V(\mathbf{p,p'})}{2 E^{(2)}_{p'}}   [(E^{(2)}_p+m_2)\boldsymbol{\sigma}\cdot \hat{\mathbf{p'}} \mathscr{Y'}_{LS}^{F m_F} \notag \\
&\;\;\;\;\;\;\;\;\;\;\;\;\;\;\;\;\;\;\;\;\;\;\;\;\;\;\;\;\;\;\;\;+S(p') p \boldsymbol{\sigma}\cdot \hat{\mathbf{p'}}\mathscr{Y'}_{LS}^{F m_F}(\boldsymbol{\sigma}\cdot \hat{\mathbf{p'}})^T (\boldsymbol{\sigma}\cdot \hat{\mathbf{p}})^T]  h_L(p')    \\
R_{22}&=-\int d^3p' \frac{V(\mathbf{p,p'})}{2 E^{(2)}_{p'}} S( p)[(E^{(2)}_p+m_2)\boldsymbol{\sigma}\cdot \hat{\mathbf{p'}} \mathscr{Y'}_{LS}^{F m_F}(\boldsymbol{\sigma}\cdot \hat{\mathbf{p}})^T \\
&\;\;\;\;\;\;\;\;\;\;\;\;\;\;\;\;\;\;\;\;\;\;\;\;\;\;\;\;\;\;\;\;\;\;\;\;\;\;\;\;+S(p') p \boldsymbol{\sigma}\cdot \hat{\mathbf{p'}} \mathscr{Y'}_{LS}^{F m_F}(\boldsymbol{\sigma}\cdot \hat{\mathbf{p'}})^T]h_L(p')
\end{align*}
where $\mathscr{Y'}_{LS}^{F m_F}=\mathscr{Y}_{LS}^{F m_F}(\Omega')$.  At this point, there is an apparent redundancy in the four equations.  Right multiplying the upper right and lower right component equations by $\boldsymbol{\sigma} \cdot \hat{\mathbf{p}}$ and dividing by $S( p)$ results in the upper left and lower left component equations.  Since we reduced the number of independent radial functions in our matrix to two by demanding that it satisfy the auxiliary condition, Eq.\eqref{eq:aux}, we expected this redundancy.  We will focus only on the left components for the remainder of this discussion. 

To keep this analysis general, we must find the action of the 
$\boldsymbol{\sigma}\cdot \mathbf{p}$ operators on the spin-angle functions,
\begin{align}
\boldsymbol{\sigma}\cdot \mathbf{p} \mathscr{Y}^{F m_F}_{LS}&=\sum_{L' S'} C^{F m_F}_{LS L' S'} \mathscr{Y}^{F m_F}_{L'S'} \\
 \mathscr{Y}^{F m_F}_{LS}(\boldsymbol{\sigma}\cdot \mathbf{p})^T&=\sum_{L' S'} {C^T}^{F m_F}_{LS L' S'} \mathscr{Y}^{F m_F}_{L'S'} 
\end{align}
where $C^{F m_F}_{LS L' S'}$ are coefficients that can be determined explicitly and tabulated.  $\boldsymbol{\sigma}\cdot \mathbf{p}$ is a pseudo-scalar operator and must change $L$ by $\pm 1$, i.e. $|L-L'|=1$.  Other properties of these coefficients are
\begin{gather}
C^{F m_F}_{L0 L' S'}=-{{C}^T}^{F m_F}_{L0 L' S'}, \notag \\
C^{F m_F}_{L1 L' 0}=-{{C}^T}^{F m_F}_{L1 L' 0}, \notag \\
C^{F m_F}_{L1 L' 1}={{C}^T}^{F m_F}_{L1 L' 1}, \notag \\
C^{F m_F}_{LS L' S'}=C^{F m_F}_{L'S' L S}, \notag \\
\sum_{L' S'}C^{F m_F}_{LS L' S'}C^{F m_F}_{L'S' L'' S''}=\delta_{L L''} \delta_{S S''}, \notag \\
\sum_{L' S'}{C^T}^{F m_F}_{LS L' S'}{C^T}^{F m_F}_{L'S' L'' S''}=\delta_{L L''} \delta_{S S''}.  \label{eq:coeff}
\end{gather}
These properties are useful when using our general equations to determine specific cases.

The partial-wave expansion of the potential is,
\begin{align}
V(\mathbf{p},\mathbf{p}')&=\frac{1}{2 \pi^2}\sum_{L=0}^\infty (2 l+1) V_L (p, p') P_L(cos \theta_{p p'}) \notag \\
&=\frac{2}{ \pi}\sum_{L=0}^\infty \sum_{m_L=-L}^L  V_L (p, p') {Y}^*_{Lm_L}(\Omega') Y_{Lm_L}(\Omega). \label{eq:partial}
\end{align}
Using the orthogonality conditions,
\begin{align}
\int_{-1}^1 dx P_{L'}(x) P_L(x)&=\frac{2}{2L+1}\delta_{L L'} \notag \\
\int d \Omega {Y}^*_{Lm_L}(\Omega) Y_{L' m_L'}(\Omega)&= \delta_{L L'} \delta_{m_L m_L'}, \label{eq:ortho}
\end{align}
the components of the partial wave expansion in terms of the potential are,
\begin{align}
V_L(p,p')= \pi^2 \int_{-1}^1 dx P_L(x) V(\mathbf{p,p'}),
\end{align}
where $x=\cos \theta_{p p'}$.  The orthogonality relation of the spin-angle functions,
\begin{align} \label{eq:ortho2}
\int d\Omega \, \text{Tr} [{\mathscr{Y}^{F m_F}_{LS}}^\dagger \mathscr{Y}_{L' S'}^{F m_F}]=\delta_{L L'},\delta_{S S'}
\end{align}
are also useful.  

Using Eq.\eqref{eq:partial} and Eq.\eqref{eq:ortho} we find the left components on the right hand side
\begin{align}
R_{11}&=\frac{2}{\pi}\int   \frac{dp' {p'}^2}{2E^{(2)}_{p'}} [(E^{(2)}_p+m_2)V_L(p,p') \mathscr{Y}_{LS}^{F m_F} \notag \\
&\;\;\;\;\;\;\;\;\;\;\;\;\;\;\;\;\;\;\;\;\;+S(p') p \sum_{L' S'}\sum_{L'' S''} {C^T}^{F m_F}_{LS L' S'}{C^T}^{F m_F}_{L'S' L'' S''} V_{L'}(p,p') \mathscr{Y}^{F m_F}_{L''S''} ] g_L(p')  \\
R_{21}&=-\frac{2}{\pi}\int \frac{dp' {p'}^2}{2E^{(2)}_{p'}}  [(E^{(2)}_p+m_2) \sum_{L' S'}{C}^{F m_F}_{LS L' S'} V_{L'}(p,p')\mathscr{Y}^{F m_F}_{L'S'} \notag \\
&\;\;\;\;+S(p') p (\sum_{L' S'} \sum_{L'' S''}\sum_{L''' S'''}{C}^{F m_F}_{LS L' S'}{C^T}^{F m_F}_{L'S' L'' S''}{C^T}^{F m_F}_{L''S'' L''' S'''} V_{L''}(p,p') \mathscr{Y}^{F m_F}_{L'''S'''}) ]  h_L(p') 
\end{align}
We remove the spin-angle functions by multiplying the top left by $(\mathscr{Y}_{LS}^{j m_j})^\dagger$ and the bottom left by $(\boldsymbol{\sigma}\cdot \hat{\mathbf{p}}\mathscr{Y}_{LS}^{j m_j})^\dagger$, taking a trace, and integrating over $\Omega$ using Eq.\eqref{eq:ortho2}.  The resulting equations are
\begin{align}
&(m_b-E^{(2)}_p-m_1) g_{L}( p)+p h_{L}( p)= \notag \\
&\;\;\;\; \frac{2}{\pi}\int \frac{dp' {p'}^2}{2E^{(2)}_{p'}} [(E^{(2)}_p+m_2)V_L(p,p')+S(p') p \sum_{L' S'} ({C^T}^{F m_F}_{LS L' S'})^2 V_{L'}(p,p') ] g_L(p') \\ 
& -p g_{L}( p)-(m_b+m_1-E^{(2)}_p)  h_L( p)= \notag \\
&\;\;\;\; -\frac{2}{\pi}\int \frac{dp' {p'}^2}{2E^{(2)}_{p'}}  [(E^{(2)}_p+m_2) \sum_{L' S'}({C}^{F m_F}_{LS L' S'})^2 V_{L'}(p,p') \notag\\
&\;\;+S(p') p \sum_{L' S'} \sum_{L'' S''}\sum_{L''' S'''}{C}^{F m_F}_{LS L' S'}{C^T}^{F m_F}_{L'S' L'' S''}{C^T}^{F m_F}_{L''S'' L''' S'''}{C}^{F m_F}_{LS L''' S'''} V_{L''}(p,p') ]  h_L(p').
\end{align}
Substituting $m_b=\epsilon+m_1+m_2$ gives
\begin{align}
\epsilon g_{L}( p)&= (E^{(2)}_p-m_2) g_{L}( p)-p h_L( p) \notag \\
&\;\;\;\;+\frac{2}{\pi}\int \frac{dp' {p'}^2}{2E^{(2)}_{p'}} [(E^{(2)}_p+m_2)V_L(p,p')+S(p') p \sum_{L' S'} ({C^T}^{F m_F}_{LS L' S'})^2 V_{L'}(p,p') ] g_L(p') \label{eq:final1} \\ 
\epsilon h_L( p)& =  (E^{(2)}_p-m_2-2m_1)  h_L( p)-p g_{L}( p) \notag \\
&\;\;\;\;+\frac{2}{\pi}\int \frac{dp' {p'}^2}{2E^{(2)}_{p'}}  [(E^{(2)}_p+m_2) \sum_{L' S'}({C}^{F m_F}_{LS L' S'})^2 V_{L'}(p,p') \notag\\
&\;\;\;\;+S(p') p \sum_{L' S'} \sum_{L'' S''}\sum_{L''' S'''}{C}^{F m_F}_{LS L' S'}{C^T}^{F m_F}_{L'S' L'' S''}{C^T}^{F m_F}_{L''S'' L''' S'''}{C}^{F m_F}_{LS L''' S'''} V_{L''}(p,p') ]  h_L(p') \label{eq:final2}
\end{align}

\subsection{Specific cases of the bound state equation}
As shown earlier, our equation reduces to the Dirac Coulomb equation in the 
large-$m_2$ limit.  Here we show this reduction for each partial wave. The last term in both equations goes to zero and the equation simplifies to
\begin{align}
\epsilon g_{L}( p)&=-p h_L( p)+\int dp'  {p'}^2 v_L(p,p') g_L( p') \\ 
\epsilon h_L( p)& =  -2m_1 h_L( p)-p g_{L}( p) +\int dp'  {p'}^2  \sum_{L' S'}({C}^{F m_F}_{LS L' S'})^2 v_{L'}(p,p') h_L( p')
\end{align}
where $v_L(p,p')=\frac{2}{\pi} V_L(p,p')$.  Again, we find the momentum space Dirac equation for an electron moving in a Coulomb potential.

We can use Eq.\eqref{eq:coeff} along with some general properties of the coefficients to simplify our equations in some specific cases.  For the case where $S=0$, $S'$ must be $1$, and we can use Eq.\eqref{eq:coeff} to greatly simplify the sums in the last term of the second equation.  The result is
\begin{align}
\epsilon g_{L}( p)&= (E^{(2)}_p-m_2) g_{L}( p)-p h_L( p) \notag \\
&\;\;\;\;+\frac{2}{\pi}\int \frac{dp' {p'}^2}{2E^{(2)}_{p'}} [(E^{(2)}_p+m_2)V_L(p,p')+S(p') p \sum_{L'} ({C^T}^{F m_F}_{L0 L' 1})^2 V_{L'}(p,p') ] g_L(p') \label{eq:simp1} \\ 
\epsilon h_L( p)& =  (E^{(2)}_p-m_2-2m_1)  h_L( p)-p g_{L}( p) \notag \\
&\;\;\;\;+\frac{2}{\pi}\int \frac{dp' {p'}^2}{2E^{(2)}_{p'}}  [(E^{(2)}_p+m_2) \sum_{L'}({C}^{F m_F}_{L0 L' 1})^2 V_{L'}(p,p') +S(p') p V_{L}(p,p') ]  h_L(p'). \label{eq:simp2}
\end{align}
For $S=1$, $L=J$, we know $L'=J\pm 1$ and $S'=1$.  Our simplified equations are
\begin{align}
\epsilon g_{L}( p)&= (E^{(2)}_p-m_2) g_{L}( p)-p h_L( p) \notag \\
&\;\;\;\;+\frac{2}{\pi}\int \frac{dp' {p'}^2}{2E^{(2)}_{p'}} [(E^{(2)}_p+m_2)V_L(p,p')+S(p') p \sum_{L'} ({C^T}^{F m_F}_{L1 L' 1})^2 V_{L'}(p,p') ] g_L(p') \\ 
\epsilon h_L( p)& =  (E^{(2)}_p-m_2-2m_1)  h_L( p)-p g_{L}( p) \notag \\
&\;\;\;\;+\frac{2}{\pi}\int \frac{dp' {p'}^2}{2E^{(2)}_{p'}}  [(E^{(2)}_p+m_2) \sum_{L' S'}({C}^{F m_F}_{L1 L' 1})^2 V_{L'}(p,p')+S(p') p  V_{L}(p,p') ]  h_L(p').
\end{align}
Finally, we have the case where $S=1$ and $L=J-1$.  In this case $L'$ must be equal to $J$, and the remaining sum of the squared coefficients over $S'$ is $1$.  The simplified equations are
\begin{align}
\epsilon g_{L}( p)&= (E^{(2)}_p-m_2) g_{L}( p)-p h_L( p) \notag \\
&\;\;\;\;+\frac{2}{\pi}\int \frac{dp' {p'}^2}{2E^{(2)}_{p'}} [(E^{(2)}_p+m_2)V_L(p,p')+S(p') p V_{J}(p,p') ] g_L(p') \label{eq:simp5} \\ 
\epsilon h_L( p)& =  (E^{(2)}_p-m_2-2m_1)  h_L( p)-p g_{L}( p) \notag \\
&\;\;\;\;+\frac{2}{\pi}\int \frac{dp' {p'}^2}{2E^{(2)}_{p'}}  [(E^{(2)}_p+m_2)  V_{J}(p,p') \notag\\
&\;\;\;\;+S(p') p \sum_{ S'} \sum_{L''}\sum_{S'''}{C}^{F m_F}_{L1 J S'}{C^T}^{F m_F}_{J S' L'' 1}{C^T}^{F m_F}_{L''1 J S'''}{C}^{F m_F}_{L1 J S'''} V_{L''}(p,p') ]  h_L(p'). \label{eq:simp6}
\end{align}

Note that even without the inclusion of a hyperfine spin-spin coupling term there is a difference between the $nS^0_0$ and the $nS^1_1$ equations.  The former's state equations, found from Eq.\eqref{eq:simp1} and Eq.\eqref{eq:simp2}, are 
\begin{align}
\epsilon g_{0}( p)&= (E^{(2)}_p-m_2) g_{0}( p)-p h_0( p) \notag \\
&\;\;\;\;+\frac{2}{\pi}\int \frac{dp' {p'}^2}{2E^{(2)}_{p'}} [(E^{(2)}_p+m_2)V_0(p,p')+S(p') p V_{1}(p,p') ] g_0(p') \\ 
\epsilon h_0( p)& =  (E^{(2)}_p-m_2-2m_1)  h_0( p)-p g_{0}( p) \notag \\
&\;\;\;\;+\frac{2}{\pi}\int \frac{dp' {p'}^2}{2E^{(2)}_{p'}}  [(E^{(2)}_p+m_2) V_{1}(p,p') +S(p') p V_{0}(p,p') ]  
h_0(p')
\end{align}
and the latter's, found from Eq.\eqref{eq:simp5} and Eq.\eqref{eq:simp6}, are
\begin{align}
\epsilon g_{0}( p)&= (E^{(2)}_p-m_2) g_{0}( p)-p h_0( p) \notag \\
&\;\;\;\;+\frac{2}{\pi}\int \frac{dp' {p'}^2}{2E^{(2)}_{p'}} [(E^{(2)}_p+m_2)V_0(p,p')+S(p') p V_{1}(p,p') ] g_0(p') \\ 
\epsilon h_0( p)& =  (E^{(2)}_p-m_2-2m_1)  h_0( p)-p g_{0}( p) \notag \\
&\;\;\;\;+\frac{2}{\pi}\int \frac{dp' {p'}^2}{2E^{(2)}_{p'}}  [(E^{(2)}_p+m_2)  V_{1}(p,p') +S(p') p (\frac{1}{9}V_{0}(p,p')+\frac{8}{9}V_{2}(p,p')) ] h_0(p') .
\end{align}
Because $p\sim \alpha \mu$, where $\mu$ is the reduced mass, the terms containing $S(p')$ are very small.  For this reason the splitting between the energy levels of these two states created by the dissimilarity in the equations is very small.  For large $m_2$, the potential terms with $S(p')$ are smaller by a factor that is $O(\alpha^2(m_1/m_2)^2)$. 

\subsection{Corrections to the approximation using the reduced mass}
The NQA introduces both the light and heavy particle masses independently, rather than introducing the heavy-particle mass via the reduced mass. In the small-p approximation, the NQA coincides with the reduced mass approximation, but for larger momenta the reduced-mass approximation fails. We show this by examining the kinetic terms in the bound-state equations,

\begin{equation}
\left[ \begin{array}{cc}
E^{(2)}_p -m_2  & -p       \\
-p                      &  E^{(2)}_p -m_2 -2m_1 \end{array} \right]
\left[\begin{array}{c}
g\\
h \end{array} \right]
=
\epsilon
\left[\begin{array}{c}
g\\
h  \end{array} \right].
\end{equation}
The eigenvalue of this equation is
\[
\epsilon=\sqrt{p^2+m_2^2}-m_2+ \sqrt{p^2+m_1^2}-m_1\approx \frac{p^2}{2 \mu} -\frac{m_1^3+m_2^3}{8m_1^3m_2^3}p^4.
\]
For the Dirac equation the kinetic terms are
\begin{equation}
\left[ \begin{array}{cc}
0  & -p       \\
-p                      &  -2\mu \end{array} \right]
\left[\begin{array}{c}
g\\
h \end{array} \right]
=
\epsilon
\left[\begin{array}{c}
g\\
h  \end{array} \right].
\end{equation}
The eigenvalue for the Dirac equation is
\[
\epsilon=\sqrt{p^2+\mu^2}-\mu \approx \frac{p^2}{2 \mu}  -\frac{p^4}{8\mu^2},
\]
where $\mu =m_1m_2/(m_1+m_2)$ is the reduced mass. (This result was found earlier by Raychaudhuri.~\cite{the}) We note that there are further mass dependencies in the potential term, but do not discuss them here.
 
\section{Numerical results}
We briefly discuss some of our numerical results here.  The purpose of this section is to show that our procedure and numerical calculations yield results that are consistent with standard calculations.  We are aware that solutions to Eq. (\ref{eq:bse3}) are already well known, and we merely intend to show that our procedure does not return any erroneous results.

We solved the integral equation numerically for several states.  We used a grid with 1200 points per equation and converted the integral eigenvalue equations into matrix eigenvalue equations.  We handled the singularities at $p=p'$ in the kernels with Lande subtractions \cite{Lande}.  We excluded momenta close to infinity to avoid infinities in our discretized integral equation.  The wave functions are extremely close to zero well before our cutoff is imposed.  We made our equations dimensionless by dividing by $m_1$ and expressed the coupled equation in terms of the dimensionless parameter $\xi=m_2/m_1$.  Our results agree with those of the Dirac-Coulomb equation with the reduced mass.  With a higher precision we expect our results to differ from the Dirac-Coulomb equation, because our equation contains effects of the proton spin that are not found in the Dirac equation.  

We found a rough estimate of our uncertainty by finding the eigenvalues with 800, 1000, and 1200 grid points and analyzing the stability of the eigenvalues.  We conservatively estimated our uncertainty to be $0.01$ meV for electronic hydrogen and $2$ meV for muonic hydrogen.  

We give comparisons of the NQA electronic hydrogen eigenvalues and Dirac-Coulomb eigenvalues for the $nS^0_0$ states in table \ref{tab:eigenvalues} \subref{subtab:subtab1}.  These values are nearly identical and the results indicate that we may have overestimated our uncertainty.  
\begin{table}[t] 
\centering
\subfloat[Electronic hydrogen $nS^0_0$]{
\begin{tabular}{| l | l | l |} \hline
    n & NQA & Dirac \\ \hline
  1 &-13.59847 & -13.59847  \\ \hline
  2 & -3.39963 & -3.39963 \\ \hline
  3 & -1.51094 & -1.51094 \\ \hline
\end{tabular} \label{subtab:subtab1}
} \;\;\;\;\;\;\;\;\;\;\;
\subfloat[Muonic hydrogen $nS^0_0$]{
\begin{tabular}{| l | l | l |} \hline
   n & NQA & Dirac \\ \hline
  1 &-2.528506 & -2.528527 \\ \hline
  2 & -0.632130 & -0.632134 \\ \hline
  3 & -0.280946 & -0.280947 \\ \hline
\end{tabular} \label{subtab:subtab2}
} \\
\caption{Energy eigenvalues for electronic and muonic hydrogen states.  The table on the left gives electronic hydrogen levels in units of eV and the right table gives muonic hydrogen levels in units of keV.  We give Dirac eigenvalues for the Dirac-Coulomb equation with the reduced mass.  We found NQA values numerically from the the NQA integral equations with a Coulomb potential. \label{tab:eigenvalues}}
\end{table}
We give the same comparisons for muonic hydrogen in table \ref{tab:eigenvalues} \subref{subtab:subtab2}.  These values are similar, but differ significantly for the lower eigenvalues.  The NQA energies in the ground- and next lowest-states are higher than the Dirac energies by $21$ meV and $4$ meV, respectively.  It is possible that our numerical calculations failed for these two particular eigenvalues, or we may have underestimated the uncertainty.  We plan on achieving a higher degree of precision in the future to investigate such concerns. 

Using the same method of estimating the uncertainty as before, we conservatively take our uncertainty to be $0.01$ meV.  As in the case of muonic hydrogen, there are some discrepancies in the first two eigenvalues.  The first and second values are larger than the Dirac energies by $0.16$ and $.03$ meV respectively.  The higher eigenvalues are consistent with the Dirac energies.

We also calculated the energies of the $nS^1_1$ states.  They are identical to those shown in tables \ref{tab:eigenvalues} \subref{subtab:subtab1} and \subref{subtab:subtab2} for some $nS^0_0$ states.  We need higher precision to study the energy splitting in these states caused by the differences in the NQA equations.  

The full coupled NQA equations are symmetric under $m_1 \leftrightarrow m_2$.  We used an approximation to get the final form of the equations used in these numerical calculations.  This approximation obscures the mass interchange symmetry, but it should still be present to some degree.  To check this, we interchanged masses and calculated a few of the eigenvalues for electronic hydrogen, where the mass interchange creates more of a drastic change to the equations than in muonic hydrogen.  We recovered the same eigenvalues shown in the tables up to 1 or 2 sigmas.  

We also found evidence that our precision is not high enough for the final terms in Eq. \eqref{eq:final1} and Eq. \eqref{eq:final2} to have a significant effect on the eigenvalues.  We calculated ground state and $n=1$ eigenvalues without these terms and the values were not appreciably different.  This is another motivation for improving our precision in the future.  

\begin{figure}[t]
\centering
\subfloat[ $1S^{0}_0$ state wave functions]{
    \label{fig:subfig1}
    \includegraphics[width=2.8in]{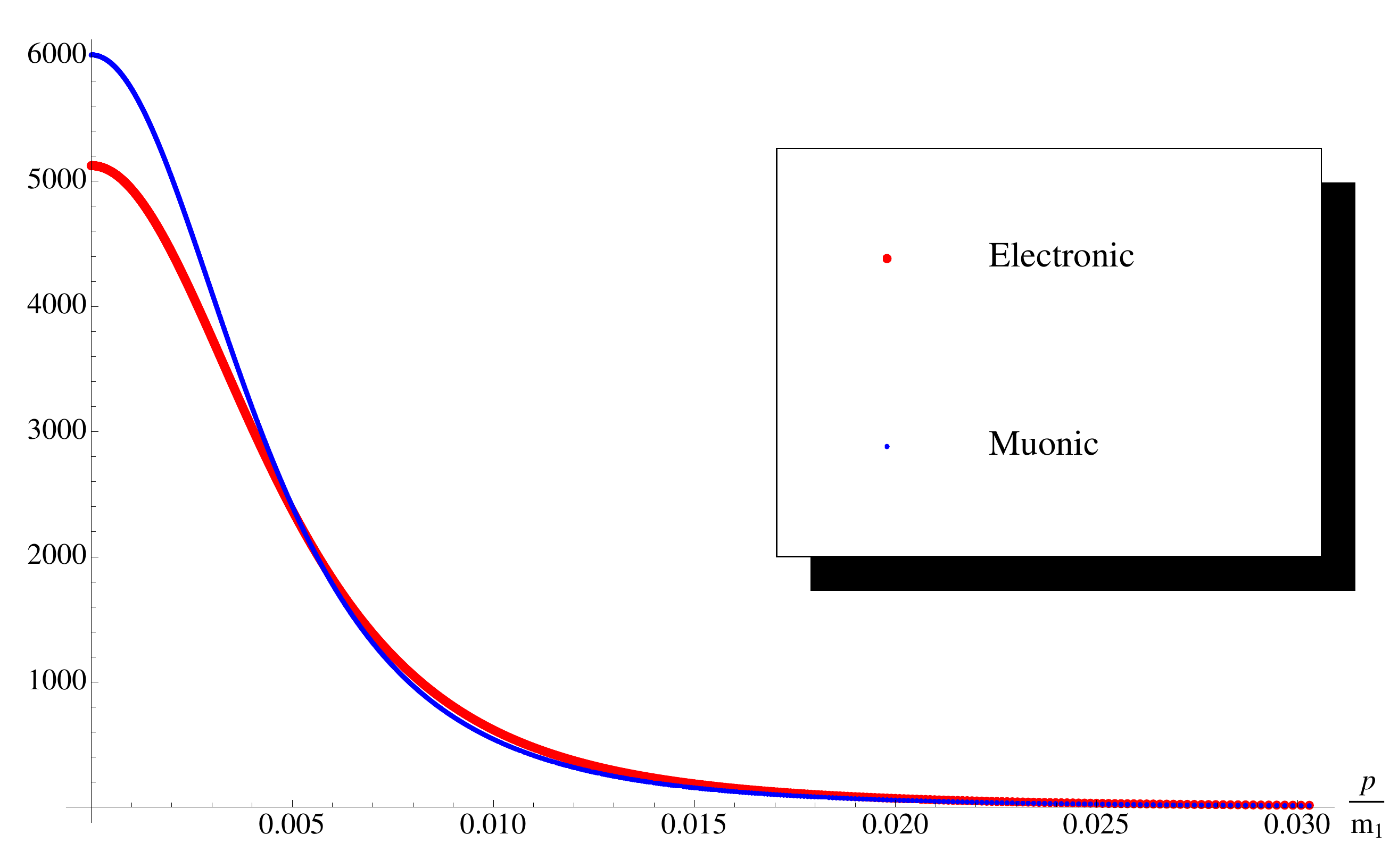}
}
\subfloat[$1S^{0}_0$ wave function differences]{
    \includegraphics[width=2.8in]{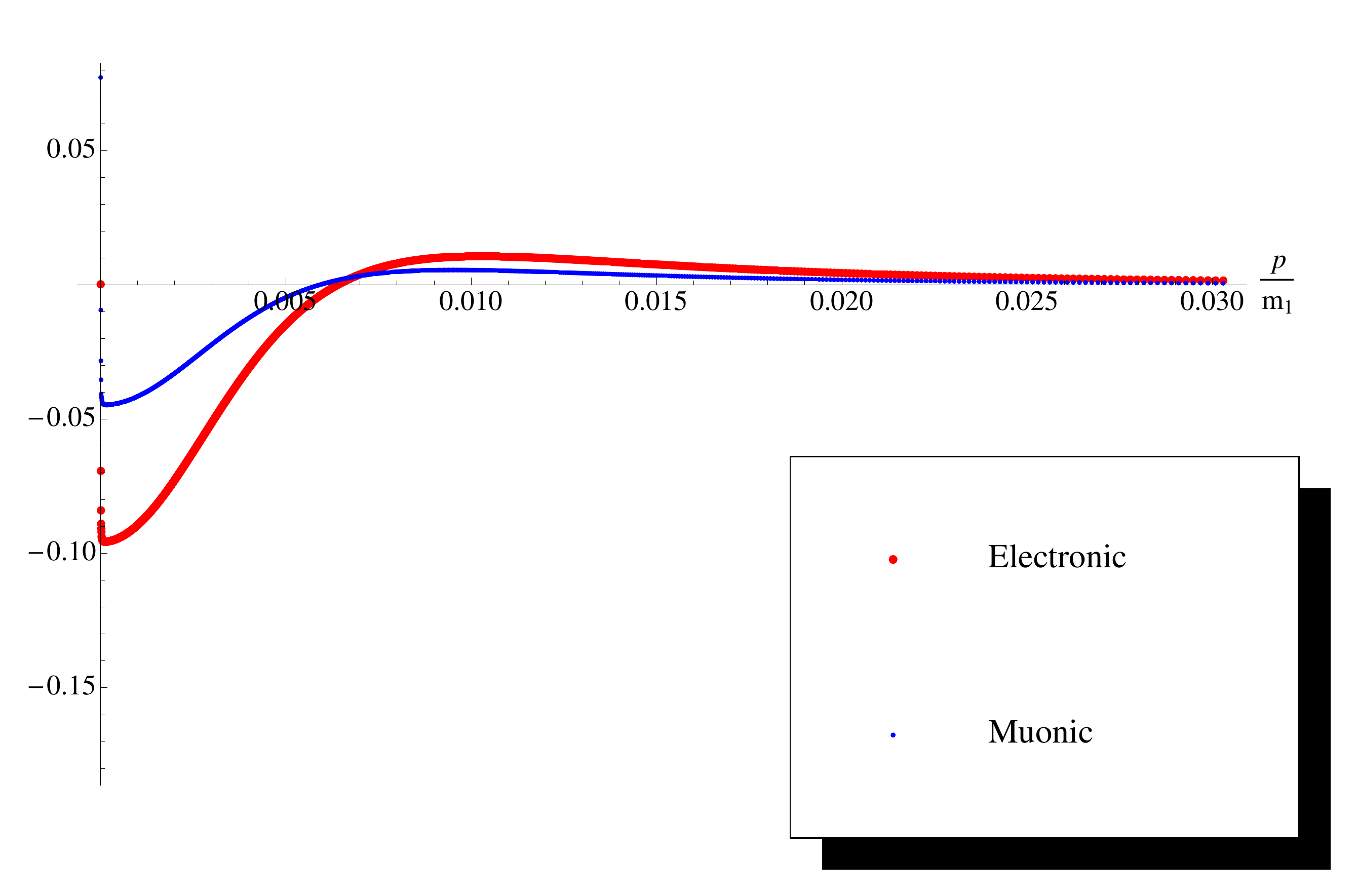}
    \label{fig:subfig2}
} \\
\subfloat[ $2S^{0}_0$ state wave functions]{
     \includegraphics[width=2.8in]{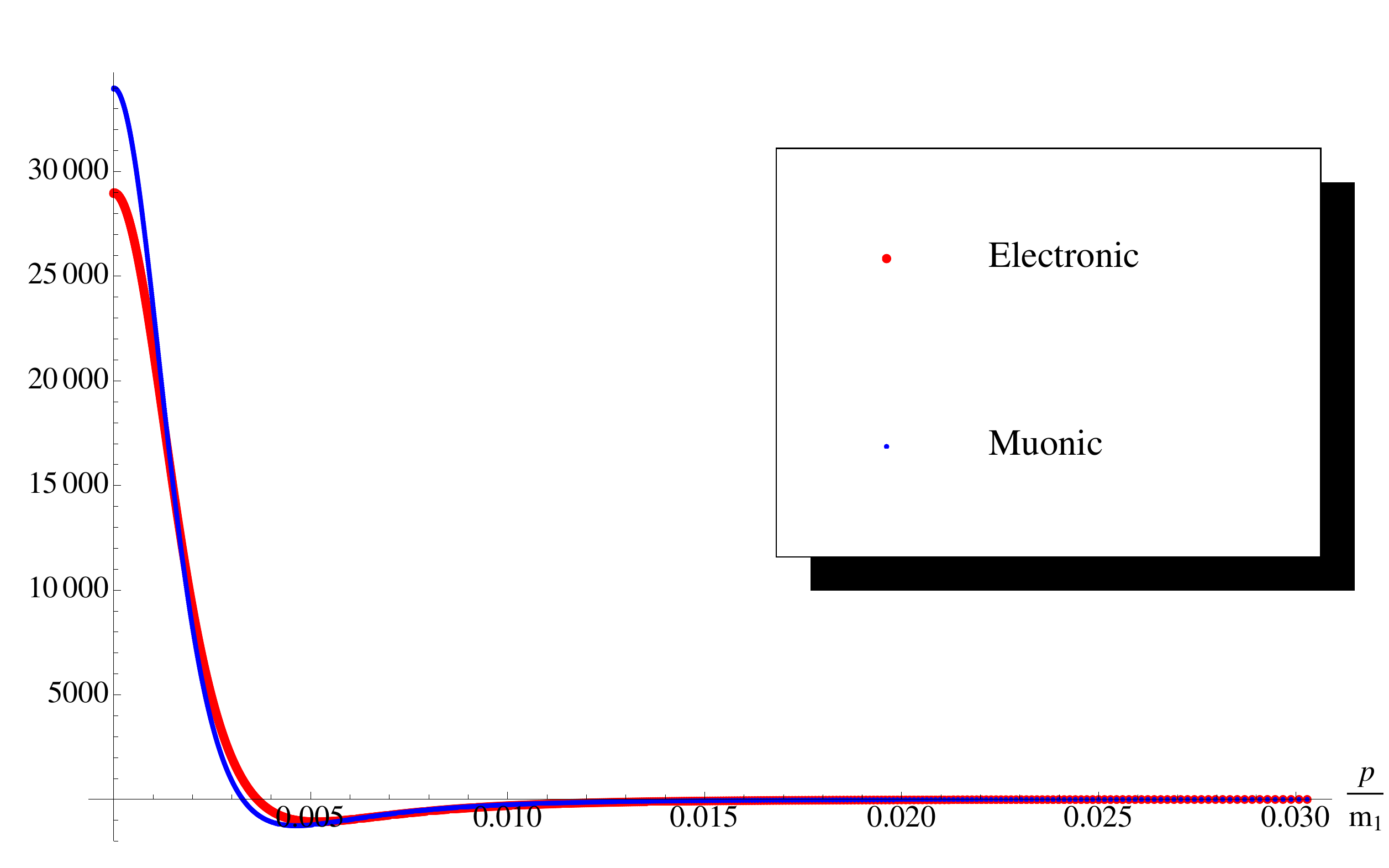}
    \label{fig:subfig3}
}
\subfloat[$2S^{0}_0$ wave function differences]{
     \includegraphics[width=2.8in]{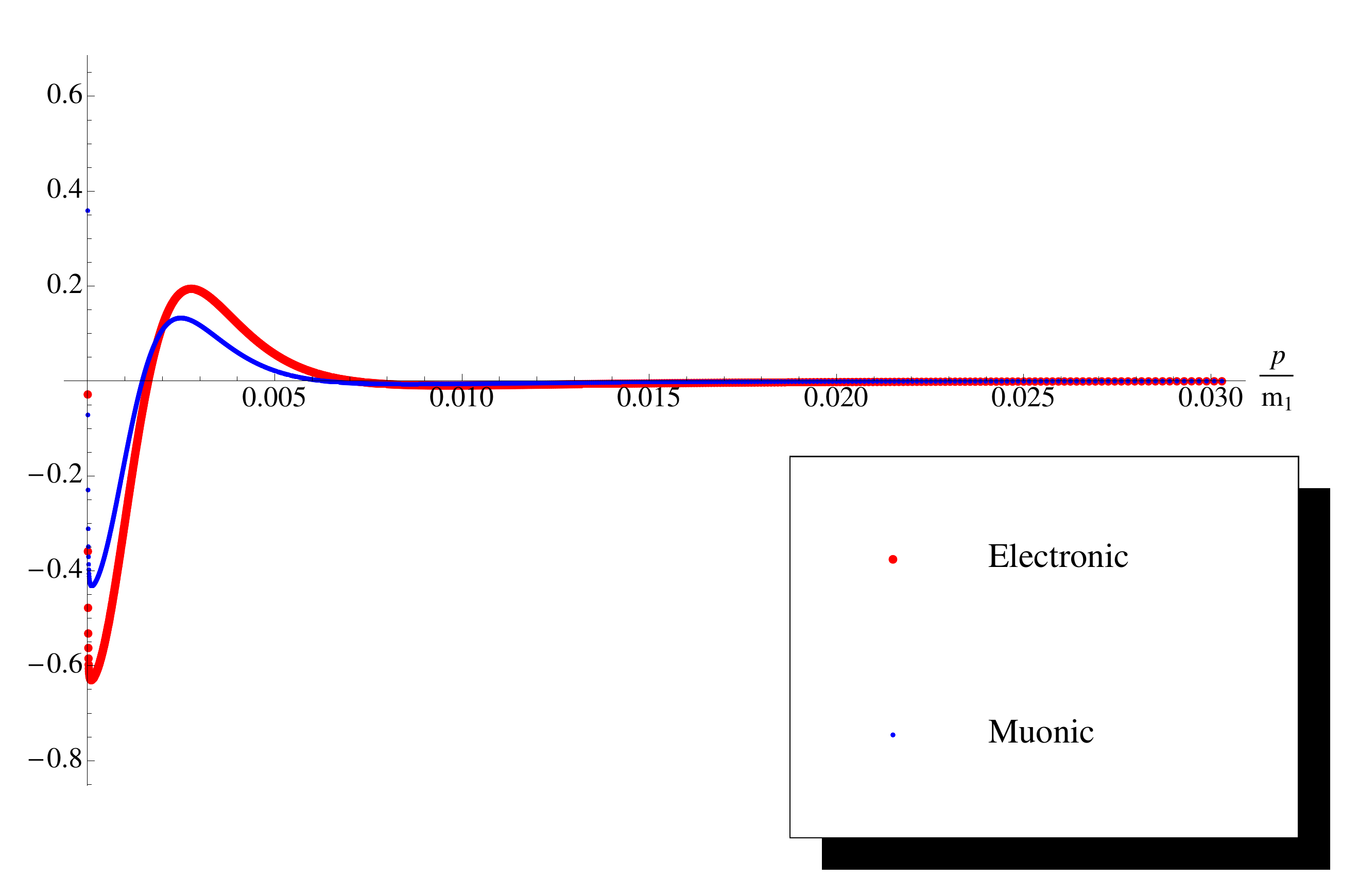}
    \label{fig:subfig4}
} \\
\caption{Plots on the left are NQA momentum space wave functions for certain electronic and muonic hydrogen states.  Plots on the right are differences between NQA wave functions and Dirac-Coulomb wave functions for certain electronic and muonic hydrogen states.  Thick red lines represent electronic hydrogen wave functions or differences and thin blue lines represent muonic hydrogen wave functions or differences.    \label{fig:wavefunctions}
}
\end{figure}

In addition to eigenvalues, we compared our wave functions with Dirac equation solutions.  
Specifically, we compared the momentum space radial wave function of the upper components of the Dirac equation solutions with our function $g_L( p)$.  Once again we make comparisons for both muonic and electronic hydrogen.  Comparisons for two states are shown in figure \ref{fig:wavefunctions}.  Plots on the left hand side are the NQA wave functions as a function of $p/m_1$ for three electronic and muonic hydrogen states.  Plots on the right hand side are differences between the NQA wave functions and Dirac-Coulomb wave functions for the same electronic and muonic hydrogen states.  We plotted the muonic and electronic wave functions and differences for each state on the same set of axes in order to make direct comparisons between electronic and muonic hydrogen.  As is the case for Dirac-Coulomb wave functions, NQA muonic wave functions are larger for smaller momenta and smaller for large momenta than their electronic counterparts.  It is clear from the right plots that muonic hydrogen wave functions are more consistent with Dirac-Coulomb wave functions.  The NQA solutions are less than the Dirac wave functions for small momenta and greater for larger momenta.

The differences shown on the right in figure \ref{fig:wavefunctions} are small compared to the size of the wave functions themselves, therefore we can conclude that we have found wave functions that are fairly consistent with Dirac wave functions, as well as eigenvalues that are all within 2 sigmas except for the $1S^0_0$ state.  It is worth noting that we introduced the heavy particle mass independently in the NQA equations, yet the results compare nicely with Dirac equation results with the reduced mass.  

\section{Framework for higher order contributions}
In this section, we show how to calculate higher order contributions to the bound state energy, specifically Lamb shift terms.  The actual calculation of these terms is beyond the scope of this paper, but we will explain how to extend our formalism to include the corrections.  The usual diagrams used to calculate the Lamb shift perturbatively are shown in figure 3.
\begin{figure}[h!]
\begin{center}
\includegraphics[scale=.75]{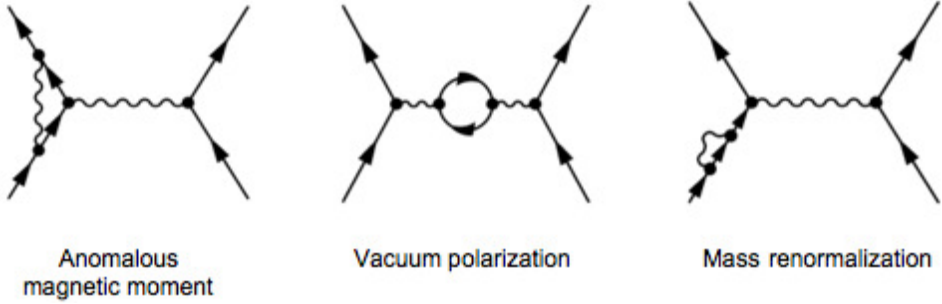}
\end{center}
\caption{Lamb shift diagrams}
\end{figure}
The energy contributions of these diagrams are typically calculated with respect to zeroth order wave functions of solutions to equations such as Eq. (\ref{eq:bse}).  We intend to calculate the analogs of these diagrams within the framework of the NQA.  We can incorporate these terms in our integral equations and use the numerical techniques described above to find a less perturbative solution.  

Three examples of NQA Lamb shift diagrams are shown in figure 4.  The external off-shell line at the lower left of each diagram is assumed to be the electron.  These terms will be added to the right hand side of Eq. (\ref{eq:bse1}).  In any diagram, every loop that does not contain an NQA amplitude must have one on-shell line.  With this rule, there are 2 permutations of the vacuum polarization diagram, 3 of the anomalous magnetic moment diagram, and 2 of the mass renormalization diagram.  The diagrams shown in figure 5 involve the amplitude $f_e$, but there are similar diagrams involving $f_p$ where the line directly to the lower left of the circle is the on-shell line and the line directly to the circle's lower right is off-shell.  This means that there are a total of 4 diagrams for vacuum polarization, 6 for the anomalous magnetic moment, and 4 for mass renormalization.  We also must find similar diagrams to add to the right hand side of Eq. (\ref{eq:bse2}) where the external off-shell line is the proton.

The NQA seems to be more complicated than the standard procedure where there is only 1 Feynman diagram for each Lamb shift contribution, but diagrams of the same type are very similar and it is not necessary to calculate each diagram explicitly.  Additionally, the mass shell delta functions that appear in the NQA simplify calculations.  
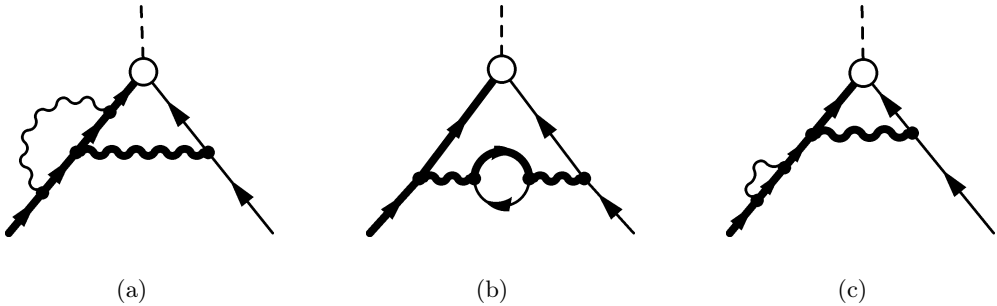
\begin{figure}[t] 
\begin{center}
\subfloat[]{\label{fig:vacb}
\begin{fmffile}{vacb}
\begin{fmfgraph*}(100,100) \fmfpen{thin}
\fmfbottomn{i}{2} \fmftop{o1} 
\fmf{fermion,tension=1.75,width=25,label=$b-p$}{i1,v1}
\fmf{photon,tension=0,left}{v1,v3}
\fmf{fermion,tension=1.75,width=25}{v1,v2}
\fmf{photon,tension=0,width=25}{v2,v4}
\fmf{fermion,tension=1.75,width=25}{v2,v3}
\fmf{fermion,tension=1.75,width=25}{v3,v5}
\fmf{fermion,label=$p$}{i2,v4}
\fmf{fermion}{v4,v5}
\fmf{dashes,tension=2,label=$b$}{v5,o1}
\fmfdot{v1,v2,v3,v4}
\fmfv{decor.shape=circle,decor.filled=empty,
decor.size=5thick}{v5}
\end{fmfgraph*}
\end{fmffile}}
\;\;\;\;\;\;
\subfloat[]{\label{fig:vaca}
\begin{fmffile}{vaca}
\begin{fmfgraph*}(100,100) \fmfpen{thin}
\fmfbottomn{i}{2} \fmftop{o1} 
\fmf{fermion,tension=2,width=25,label=$b-p$}{i1,v1}
\fmf{fermion,width=25}{v1,v5}
\fmf{photon,width=25,tension=.3}{v1,v2}
\fmf{fermion,left,width=25,tension=.15}{v2,v3}
\fmf{fermion,left,tension=.15}{v3,v2}
\fmf{photon,width=25,tension=.3}{v3,v4}
\fmf{fermion,tension=2,label=$p$}{i2,v4}
\fmf{fermion}{v4,v5}
\fmf{dashes,tension=3,label=$b$}{v5,o1}
\fmfdot{v1,v2,v3,v4}
\fmfv{decor.shape=circle,decor.filled=empty,
decor.size=5thick}{v5}
\end{fmfgraph*}
\end{fmffile}}
\;\;\;\;\;\;
\subfloat[]{\label{fig:vacc}
\begin{fmffile}{vacc}
\begin{fmfgraph*}(100,100) \fmfpen{thin}
\fmfbottomn{i}{2} \fmftop{o1} 
\fmf{fermion,tension=4.25,width=25,label=$b-p$}{i1,v1}
\fmf{photon,tension=0,left}{v1,v2}
\fmf{fermion,tension=4.25,width=25}{v1,v2}
\fmf{fermion,tension=4.25,width=25}{v2,v3}
\fmf{photon,tension=0,width=25}{v3,v4}
\fmf{fermion,tension=2.25,width=25}{v3,v5}
\fmf{fermion,label=$p$,tension=1.5}{i2,v4}
\fmf{fermion,tension=2.5}{v4,v5}
\fmf{dashes,tension=3.75,label=$b$}{v5,o1}
\fmfdot{v1,v2,v3,v4}
\fmfv{decor.shape=circle,decor.filled=empty,
decor.size=5thick}{v5}
\end{fmfgraph*}
\end{fmffile}}

\end{center}
\caption{Examples of NQA Lamb shift diagrams}
\end{figure}

\section{Summary and future work}
We used the N-Quantum Approach (NQA) in one-loop order to calculate the 
energy levels and bound-state amplitudes of ordinary and muonic hydrogen 
and of positronium. We can treat other two-body systems, such as the ($e 
\bar{\mu}$) system and the ($\mu \bar{\mu}$) system, in an analogous 
way. We used the NQA systematically to find a relation between wave functions with the light particle off-shell and the heavy particle off-shell and to find normalization conditions for our amplitudes.
 
We will use perturbation theory to add corrections to the electron-photon vertex and to the photon propagator to include the terms that lead to the Lamb shift.  
 
In future work we will derive integral equations that include 
higher-order corrections. We will solve these equations numerically 
without using perturbation theory in order to include some of the energy correction terms that are usually calculated perturbatively. We will compute the Lamb shift and the hyperfine structure of both electronic and muonic hydrogen. We plan to carry these calculations to sufficient order to compare our 
results with the usual methods to see if our methods resolve the 
muonic hydrogen anomaly.  

\flushleft
{\Large{\textbf{Acknowledgements}}}\\
~
This work was supported in part by the Maryland Center for Fundamental Physics. OWG thanks Professor Nicola Khuri for his hospitality at the Rockefeller University where part of this work was carried out. We are especially grateful to Steve Wallace for an extremely careful reading of the manuscript and for many very helpful suggestions.

\end{document}